\documentclass[prl,nofootinbib,twocolumn,superscriptaddress]{revtex4}
\usepackage{amsmath}

\newcommand{\G}{{\cal G}^{\rm SM}}
\renewcommand{\H}{{\cal H}^{\rm SM}}

\renewcommand{\t}{\tilde t}
\renewcommand{\u}{\tilde u^{(2)}}
\renewcommand{\b}{\tilde b}
\renewcommand{\d}{\tilde d^{(2)}}

\newcommand{\ubar}{\overline{\tilde u}^{(2)}}
\newcommand{\ubarL}{\overline{\tilde u^{(2)}_L}}
\newcommand{\bbar}{\overline {\tilde b}}

\newcommand{\dbarL}{\overline{\tilde d^{(2)}_L}}
\newcommand{\dbarR}{\overline{\tilde d^{(2)}_R}}

\newcommand{\Lgmfv}{\Lambda_{\rm MFV}}

\newcommand{\be}{\begin{equation}}
\newcommand{\ee}{\end{equation}}
\newcommand{\bea}{\begin{eqnarray}}
\newcommand{\eea}{\end{eqnarray}}
\newcommand{\beq}{\begin{equation}}
\newcommand{\eeq}{\end{equation}}
\newcommand{\beqa}{\begin{eqnarray}}
\newcommand{\eeqa}{\end{eqnarray}}

\renewcommand{\Im}{{\cal I}m}
\newcommand{\diag}{{\rm diag\,}}

\begin{document}
%\baselineskip=18pt
%\linespread{1.3}
%\begin{titlepage}
%\begin{center}

 \title{General Minimal Flavor Violation}
  
 \author{Alexander L. Kagan} \address{Department of Physics, University
   of Cincinnati, Cincinnati, Ohio 45221, USA}
  
 \author{Gilad Perez} \address{Department of Particle Physics, Weizmann
   Institute of Science, Rehovot 76100, Israel}
   \address{YITP, Stony Brook University, Stony Brook, NY 11794-3840, USA }

 \author{Tomer Volansky} \address{School of Natural Sciences, Institute
   for Advanced Study, Princeton, NJ 08540}

 \author{ Jure Zupan} \address{Theory Division, Department of Physics,
   CERN, CH-1211 Geneva 23, Switzerland}  \address{Faculty of mathematics
   and physics, University of Ljubljana, Jadranska 19, 1000 Ljubljana,
   Slovenia} \vskip .25in

 \begin{abstract}
 A model independent study of the minimal flavor violation (MFV)
   framework is presented, where the only sources of flavor breaking at low energy are
   the up and down Yukawa matrices.  Two limits are identified for the Yukawa coupling expansion:  linear MFV, where it is truncated at the leading terms, and nonlinear MFV, where such a truncation is not possible due to large third generation Yukawa couplings. 
These are then resummed to all orders using non-linear $\sigma$-model techniques familiar from models of collective breaking. Generically, flavor diagonal CP violating (CPV) sources in the UV can induce $O(1)$ CPV in processes involving third generation quarks.
Due to a residual U(2) symmetry, the extra CPV in $B_d-\bar B_d$ mixing
is bounded by CPV in $B_s-\bar B_s$ mixing. If operators with right-handed light quarks are subdominant, the extra CPV is equal in the two systems, and is negligible in processes involving only the first two generations.  We find large enhancements in the up type sector,
both in CPV in $D-\bar D$ mixing and in top flavor
violation.

 \end{abstract}

 \vskip .2in

 \maketitle

{\bf Introduction. } Precision flavor and CP
violation measurements
provide very strong constraints on models of new
physics (NP) beyond the Standard Model (SM).
For instance, $\epsilon_K$ constrains
the scale of maximally flavor violating NP to be $\gtrsim 10^4$ TeV.
Therefore, TeV scale NP which stabilizes the electroweak scale and is accessible at the LHC has to have a highly non generic flavor structure. 

The tension with precision flavor  tests is relaxed if the SM Yukawa matrices are the only
source of flavor breaking, even in the presence of new particles and
interactions~\cite{Hall:1990ac,D'Ambrosio:2002ex,Buras}. This hypothesis goes under the name of Minimal Flavor Violation (MFV). Sometimes additional assumptions are made --- that the SM Yukawa couplings are also the only source of CP violation (CPV ), {\it e.g.} in~\cite{D'Ambrosio:2002ex}, or that NP does not change the Lorentz structure of  the effective weak hamiltonian~\cite{Buras:2003jf}. We will not make these assumptions, but will discuss their consequences below.

A useful language for discussing MFV was introduced in~\cite{D'Ambrosio:2002ex}. It relies on the observation that for vanishing Yukawa couplings the SM has an enhanced global symmetry. Focusing on the quark sector this is
\begin{eqnarray}
  \label{eq:20}
  \G = U(3)_Q\times U(3)_u\times U(3)_d,
\end{eqnarray}
where $Q,u,d$ stand for quark doublets and up and down type quark singlets respectively. The SM Yukawa couplings
\begin{eqnarray}
  \label{eq:2}
   H_u\bar Q_L Y_u u_R+H_d \bar Q_LY_d d_R,
\end{eqnarray}
are formally invariant under $\G$, if the Yukawa matrices are promoted to spurions that
transform as $
   Y_{u,d}'=V_{Q} Y_{u,d} V_{u,d}^\dagger,$
while the quark fields are in the fundamental representations, $(Q',u',d')=V_{Q,u,d} (Q,u,d)$. Weak scale NP models are then of the MFV class if they are formally invariant under $\G$, when treating the SM Yukawa couplings  as spurions. Similarly, the low energy flavor observables are formally invariant under $\G$. Practically, this means that only certain insertions of Yukawa couplings are allowed in the quark bilinears. For example, in $\bar Q Q$ bilinears insertions such as $\bar Q (Y_u Y_u^\dagger )^n Q$ are allowed, while $\bar Q Y_d^\dagger (Y_u Y_u^\dagger)^n Q$ are not.

The above definition of MFV is only useful if flavor invariant operators such as $\bar Q f(\epsilon_u Y_u, \epsilon_d Y_d)Q$ can be expanded in powers of $Y_{u,d}$.  In the large $\tan \beta$ limit both $Y_{u}$ and $Y_{d}$ have $O(1)$ eigenvalues $y_{t,b}$. The convergence radius is then given by the size of $\epsilon_{u,d}$.  We distinguish between two limiting cases
\begin{itemize}
\item 
{\sl Linear
  MFV (LMFV)}:  $\epsilon_{u,d}\ll 1$ and the dominant flavor breaking effects are captured by the lowest order polynomials of $Y_{u,d}$.   
\item 
{\sl Non-linear MFV (NLMFV)}:  $\epsilon_{u,d}\sim O(1)$,  higher powers of $Y_{u,d}$ are important, and a truncated expansion in  $y_{t,b}$ is not possible.
\end{itemize}
Examples of NLMFV are: low energy supersymmetric models in which large $\tan\beta$ effects need to be resummed (large $\epsilon_d$), and models obeying MFV at a UV scale $\Lambda_F\gg \mu_W$, where large $\epsilon_{u,d} \propto \log(\mu_W/\Lambda_F)$ are generated from sizable anomalous dimensions in the renormalization group running~\cite{Paradisi:2008qh}. Another example is warped extra dimension models with alignment~\cite{Fitzpatrick:2007sa}, in cases where right handed up-quark currents are subdominant.

In this letter we show that even in NLMFV there is a systematic expansion in small quantities, $V_{td},V_{ts}$, and light quark masses, while resumming in $y_t, y_b\sim O(1)$. 
This is achieved via a non-linear  $\sigma$-model--like parametrization.  Namely, in the limit  of vanishing weak gauge coupling  (or $m_W\to \infty$), $U(3)_Q$ is enhanced to $U(3)_{Q^u}\times U(3)_{Q^d}$.  The two groups are broken down to $U(2)\times U(1)$ by large third generation eigenvalues in $Y_{u,d}Y_{u,d}^\dagger$, so that the low energy theory is described by a $[U(3)/U(2)\times U(1)]^2$ non-linear $\sigma$-model.  Flavor violation arises due to the misalignment of $Y_u$ and $Y_d$, given by $V_{td}$ and $V_{ts}$ once the weak interaction is turned on. 
We can then prove with complete generality that in MFV: (i) extra CPV can only arise from flavor diagonal CPV sources in the UV theory;
(ii)  the extra CP phases in $B_s-\bar B_s$ mixing provide an upper bound on the amount of CPV in $B_d-\bar B_d$ mixing;
(iii) if operators containing right-handed light quarks are subdominant then the extra CPV is equal in the two systems,
and is negligible in $2\to1$ transitions.
Conversely, these operators can break the correlation between CPV
in the $B_s$ and $B_d$ systems, and can induce significant new CPV in $\epsilon_K$.
%Furthermore, if all effective operators are roughly suppressed by the same scale then the extra CPV in the two systems is equal,
%and  it is parametrically suppressed in $2\to1$ transitions. 
Combinations of observables which are sensitive to LMFV vs. NLMFV are also identified.
Another non-linear parameterization of MFV was presented in~\cite{Feldmann:2008ja}.  We focus on exploiting the general control
obtained by our formalism in order to study its model independent implications.
 A modification of the formalism is needed for $y_b\ll 1$, as discussed below.

{\bf Formalism.} To realize $\G$ non-linearly, we promote the Yukawa matrices  to spurions, with the transformation properties given below Eq. \eqref{eq:2}. These flavor transformations are
broken once the Yukawa couplings obtain their background values. The
eigenvalues of the latter are hierarchical and the two matrices are
approximately aligned. We therefore take $Y_u\sim \diag(0,0,y_t)$ and $Y_d\sim \diag(0,0,y_b)$.  The
breaking of the flavor group is dominated by the top and bottom Yukawa
couplings which break it down to  $\H=U(2)_{Q}\times U(2)_{u}\times U(2)_{d} \times U(1)_3$.

The broken symmetry generators live in $\G/\H$ cosets. 
It is useful to factor them out of the Yukawa matrices. We thus use the parameterization 
\begin{eqnarray}
  \label{eq:3}
  \label{Yukawa-decomp} Y_{u,d}=e^{i \hat \rho_Q} e^{\pm i \hat
  \chi /2} \tilde Y_{u,d} e^{-i\hat \rho_{u,d}},
\end{eqnarray}
where the reduced Yukawa
spurions, $ \tilde Y_{u,d} $, are
\begin{eqnarray}
  \label{eq:4}
  \tilde Y_{u,d}= \begin{pmatrix}
  \phi_{u,d} & 0\\
  0 & y_{t,b}
\end{pmatrix}.
\end{eqnarray}
Here $\phi_{u,d}$ are $2\times 2$ complex spurions, while $\hat\chi$
and $\hat \rho_i$, $i=Q,u,d$, are the $3\times 3$ matrices spanned by the
broken generators. Explicitly,
\begin{eqnarray}
  \label{eq:5}
  \hat \chi=
  \begin{pmatrix}
    0 & \chi\\
    \chi^\dagger & 0
  \end{pmatrix},
  \qquad
  \hat \rho_{i}=
  \begin{pmatrix}
    0 & \rho_{i}\\
    \rho_{i}^\dagger & \theta_{i} %JZ corrected k->i
\end{pmatrix},\qquad
i=Q,u,d,
\end{eqnarray}
where $\chi$ and $\rho_i$ are two dimensional vectors.  The $\rho_i$ shift under
the broken generators and therefore play the role of spurion
"Goldstone bosons".  Thus the $\rho_i$ have no physical significance.   $\chi$, on the other hand, parametrizes the misalignment of the up and down Yukawa couplings and will therefore correspond to $V_{td}$ and $V_{ts}$ in the low energy effective theory [see Eq.~(\ref{eq:chi})].

Under the
flavor group the above spurions transform as, 
\begin{eqnarray}
  \label{eq:9}
   e^{i\hat\rho_{i}'} = V_ie^{i\hat \rho_{i}}U_i^\dagger, \ \
   e^{i\hat\chi'} = U_Q e^{i\hat\chi} U_Q^\dagger, \ \
   \tilde Y_{i}^\prime = U_Q \tilde Y_{i} U_{i}^\dagger.
\end{eqnarray}
Here $U_{i}=U_{i}(V_i,\hat \rho_i)$ are (reducible) unitary
representations of the unbroken flavor subgroup $U(2)_i\times U(1)_3$,
\begin{eqnarray}
  \label{eq:6}
  U_{i}=
  \begin{pmatrix}
    U_{i}^{2\times2} & 0\\
    0 & e^{i\varphi_3}
  \end{pmatrix},
~~i=Q,u,d. 
\end{eqnarray}
For $V_i\in\H$, $U_i=V_i$.  Otherwise the $U_i$ depend on the broken
generators and $\hat\rho_i$.  They form a nonlinear realization of the full flavor group. In particular, Eq. \eqref{eq:9} defines
$U_{i}(V_i,\hat \rho_i)$ by requiring that $\hat \rho_i'$ is of the
same form as $\hat \rho_i$, Eq.~(\ref{eq:5}).  Consequently $\hat\rho_i$ is shifted under $\G/\H$ and can be set to a convenient value as discussed below.  Under $\H$,  $\chi$\,[$\rho_i$] are
fundamentals of $U(2)_Q$\,[$U(2)_i$] carrying  charge $-1$ under the
$U(1)_3$, while $\phi_{u,d}$ are bi-fundamentals of $U(2)_Q\times
U(2)_{u,d}$.  

As a final step we also redefine the quark fields by 
moding out the "Goldstone spurions",
\begin{eqnarray}
  \label{def-tilded-q}
  \tilde u_L= e^{-i \hat \chi /2} e^{-i \hat \rho_Q}  u_L, &\quad \tilde d_L=
  e^{ i \hat \chi /2} e^{- i \hat \rho_Q} d_L,
  \\ 
  \label{eq:7}
  \tilde u_R= e^{-i \hat \rho_u} u_R, \qquad  & \tilde d_R= e^{-i \hat \rho_d} d_R.
\end{eqnarray}
The latter 
form reducible representations of $\H$.  Concentrating here and below on the down sector
  we therefore define  $\tilde d_{L,R}=(\d_{L,R},0)+(0,\b_{L,R})$. 
   Under flavor transformations $\d_{L}{}'=U_{Q}^{2\times2} \d_{L}$ and $\b_L{}'=\exp(i \varphi_3)\b_L$.  A similar definition can be made for the up quarks.

With the redefinitions above, invariance under the full flavor group
is captured by the invariance under the unbroken flavor subgroup
$\H$~\cite{Weinberg}.  
\emph{Thus, {\rm NLMFV} can be described without loss of generality as a formally
  $\H$--invariant expansion in $\phi_{u,d}, \chi$.}
This is a straightforward generalization of the known
effective field theory description of 
spontaneous symmetry
breaking~\cite{Weinberg}.  The only difference in our case is that
$Y_{u,d}$ are not aligned, as 
manifested by $\chi\neq0$. Since the background field values of the
relevant spurions are small, we can expand in them. 

We are now in a position to write down the flavor structures of quark
bilinears from which low energy flavor observables can be constructed. We work to leading order in the spurions that break
$\H$, but to all orders in the top and bottom Yukawa couplings.  Beginning with
the left-left (LL) bilinears, to
second order in $\chi, \phi_{u,d}$ one finds (omitting  gauge and Lorentz indices)
\begin{eqnarray}
  \label{eq:8}
  &  \bbar_L \b_L, \quad \dbarL\d_L, \quad \dbarL\phi_{u}
  \phi_{u}^\dagger \d_L,
  \\
  \label{eq:10}
  &  \dbarL\chi \b_L,\quad \bbar_L\chi^\dagger \chi \b_L,\quad \dbarL\chi \chi^\dagger
  \d_L. 
\end{eqnarray}
The first two bilinears in Eq.~\eqref{eq:8} are diagonal in the down-quark
mass basis and do not induce flavor violation.  In this basis the Yukawa couplings  take the form $Y_u=V_{\rm CKM}^\dagger \diag(m_u,m_c,m_t),\,Y_d=\diag(m_d,m_s,m_b)$. This corresponds to spurions taking the background values $\rho_Q = \chi  /2$, $\hat \rho_{u,d}=0$,
 $\phi_d=\diag(m_d,
m_s)/m_b$, while flavor violation is induced via 
\beq
\label{eq:chi}
 \chi^\dagger= i(V_{td},V_{ts}),\qquad \phi_u=V_{\rm CKM}^{(2)\dagger}\,\diag\big(\frac{m_u}{m_t},\frac{m_c}{m_t}\big).
\eeq
$V_{\rm CKM}^{(2)}$ stands for a two generation CKM matrix.
In terms of $\lambda=\sin\theta_C\simeq 0.23$, the flavor violating spurions scale as  
 $  \chi  \sim (\lambda^3, \lambda^2)$, $(\phi_u)_{12} \sim \lambda^5$.
Note that the 
redefined down quark fields, Eqs.~(\ref{def-tilded-q},\ref{eq:7}), coincide
with the mass-eigenstate basis, $\tilde
d_{L,R}=d_{L,R}$, for the above choice of spurion background values.  

The left-right (LR) and right-right (RR) bilinears which contribute to flavor mixing are in turn (at leading order in $\chi,\phi_{u,d}$ spurions), %JZ at third order the first operator
%would have a correction, so stating that these are all up to fourth order was not correc
\begin{eqnarray}
  \label{eq:13}
  &
 \dbarL \chi \b_R,  \quad  \dbarL\chi\chi^\dagger \phi_d \d_R, \quad  \bbar_L \chi^\dagger \phi_d \d_R,
  \\
  \label{eq:13a}
  &
   \dbarR \phi_d^\dagger \chi  \b_R, \quad \dbarR\phi_d^\dagger \chi \chi^\dagger \phi_d \d_R.
\end{eqnarray}

To make contact with the more familiar MFV notation, consider down quark flavor violation from LL bilinears. 
We can then expand in the  Yukawa couplings, 
\beq
\label{eq:a1}
\bar Q \big[a_1 Y_{u}Y_{u}^\dagger +a_2 (Y_{u}Y_{u}^\dagger)^2 \big]Q+ \big[ b_2\, \bar Q Y_{u}Y_{u}^\dagger Y_{d}Y_{d}^\dagger Q +h.c.\big]+\cdots,
\eeq
with $a_{1,2}= O(\epsilon_u^{2,4})$, $b_2= O(\epsilon_u^2 \epsilon_d^2 )$. Following the discussion in the Introduction, the LMFV limit corresponds to $a_1\gg a_2,b_2$, and the NLMFV limit to $a_1\sim a_2\sim b_2$. 
While $a_{1,2}$ are real, the third 
 operator in Eq. \eqref{eq:a1} is not Hermitian and 
$b_2$ can be complex~\cite{Colangelo:2008qp}, introducing a new CP violating phase beyond the SM phase. The leading flavor violating terms in Eq. \eqref{eq:a1} for the down quarks 
are
\beqa
\label{compareLMFV}
 &&\hspace*{-.8cm}
\bar d_L^i \big[(a_1+a_2 y_t^2)\xi^t_{ij}+a_1\xi^c_{ij}\big] d_L^j
+ 
\big[ b_2 y_b^2 \,  \bar d_L^i \xi^t_{ib} b_L+ h.c.\big] 
=\nonumber\\ 
&&\hspace*{-.8cm} c_b \big(\dbarL \chi \b_L+h.c\big)+
c_t \dbarL\chi \chi^\dagger \d_L+ c_c\dbarL\phi_{u}
  \phi_{u}^\dagger \d_L\,,
\eeqa
where $\xi^k_{ij}=y_k^2 V_{ki}^* V_{kj}$ with $i\neq j$.  On the RHS we have used the general parameterization in Eqs. (\ref{eq:8},\ref{eq:10}) with $c_b\simeq(a_1 y_t^2+a_2 y_t^4+b_2 y_b^2),\, c_t\simeq a_1 y_t^2+a_2 y_t^4$
and $c_c\simeq a_1 $ to leading order.  The contribution of the $c_c$ bilinear in flavor changing transitions is $O(1\%)$ compared to the $c_t$ bilinear, and can be neglected in practice. 

{\bf LMFV vs. NLMFV.} 
A novel feature of NLMFV is the potential for observable CPV from right-handed currents, to which we return below.
Other important distinctions can be readily understood from 
Eq. (\ref{compareLMFV}).  In NLMFV (with large $\tan\beta$) the extra flavor diagonal CPV phase $\Im(c_b)$ can be large, leading to observable deviations in the $B_{d,s}-\bar B_{d,s}$ mixing phases, but none in LMFV. 
%Since $c_{c,t}$ are real, this extra CPV would not affect the kaon system.
Another example is $b\to s \nu \bar{\nu}$ and $s \to d \nu \bar{\nu}$ transitions. These   
receive contributions only from a single operator in Eq. (\ref{compareLMFV}) multiplied by
the neutrino currents.  Thus, new contributions to $B \to X_s \nu \bar{\nu}$, $B \to K \nu \bar{\nu}$ vs. $K_L \to \pi^0 \nu\bar{\nu}$, $K^+ \to \pi^+ \nu \bar{\nu}$ 
are correlated in LMFV ($c_b \simeq c_t $), see e.g.,~\cite{Bobeth:2005ck}, but are independent 
in NLMFV with large $\tan \beta$.
$O(1) $ effects in the rates would correspond to 
an effective scale $\Lambda_{\rm MFV} \sim 3$ TeV in the four fermion operators, 
with smaller effects scaling like $1/\Lambda_{\rm MFV}$ due to interference with the SM contributions. 
Other interesting NLMFV effects involving the third generation, e.g., large deviations
in ${\rm Br}(B_{d,s}\rightarrow \mu^+\mu^-)$ and $b\rightarrow s\gamma$, arise in the MSSM at
large $\tan\beta$, where resummation is required~\cite{Bobeth:2002ch}. 
Contributions to $1\to2$ transitions which proceed through the charm ($c_c$) and the top ($c_t$)
are correlated within LMFV ($c_t\simeq c_c y_t^2$), but are independent in the NLMFV case, even for small $\tan\beta$. Unfortunately, the smallness of the $c_c$ bilinear prevents tests of this correlation in the near future, e.g., 
via comparison of $K^+\to \pi^+ \nu\bar\nu$ and  the CPV decay $K_L\to \pi^0 \nu\bar\nu$.

%is only sensitive to the  CP violating part of $c_t$.  Unfortunately, the smallness of the $c_c$ bilinear prevents tests of this correlation in the near future.

{\bf CP Violation.}
%The phenomenological consequences of the LMFV hypothesis, with the additional assumption of CP broken only by SM Yukawa couplings, was analyzed in detail in the literature (see {\it e.g.}~\cite{Buras:2003jf} and Refs. therein.).  
%Here we focus on the case where new CP violating phases exist, and on the phenomenological implications of the NLMFV limit. 
Assuming MFV, new CPV effects can be significant if and only if the UV theory contains new flavor-diagonal CP sources.
%and
%(ii) At leading order they appear only in $3\to 1$ and $3\to 2$ transitions.
The proof is as follows.  
If no flavor diagonal phases are present, CPV only arises from the CKM phase. In the exact 
$U(2)_L$ limit the CKM phase can be removed and the theory becomes CP invariant (at all scales). The only spurions that break the $U(2)_L$ flavor symmetry are $\phi_{u,d}$ and $\chi$. 
CPV in operators linear in $\chi$ is directly proportional to the CKM phase [cf. Eq. \eqref{compareLMFV}]. Any additional contributions are suppressed by at least $[\phi_u^\dagger \phi_u, \phi_d^\dagger \phi_d]\sim (m_s/m_b)^2 (m_c/m_t)^2 \sin\theta_C\sim 10^{-9}$, and are therefore negligible.

Flavor diagonal weak phases in NLMFV can lead to new CPV effects in $3\to 1$ and $3\to 2$ decays.
An example is $\Delta B=1$ electromagnetic and chromomagnetic dipole operators
constructed from the first bilinear in Eq. \eqref{eq:13}. The operators are not Hermitian,
hence their Wilson coefficients can
contain new CPV phases.
Without new phases,
the untagged direct CP asymmetry in $B \to X_{d,s} \gamma$
would essentially vanish due to the residual $U(2)$ symmetry, as in the SM~\cite{soares}, and the $B \to X_s \gamma$ 
asymmetry would be less than a percent.
However, in the NLMFV limit (large $y_b$), non-vanishing phases can yield significant CPV in untagged and
$B \to X_s \gamma$ decays, and the new CPV in $B \to X_s \gamma$ and $B \to X_d \gamma$ would be strongly correlated.
Supersymmetric examples of this kind were studied in~\cite{Hurth:2003dk}, where new phases were discussed.

Next, consider the NLMFV $\Delta B=2$ effective operators. They are not
Hermitian, hence  their Wilson coefficients $\kappa_i /\Lgmfv^2$ can 
also contain new CP violating phases.
The operators can be divided into two classes:  class-1, which do not
contain light right-handed quarks [$(\overline{\d_L}\chi\b_{L,R})^2 $,...]; and class-2, which do
[$(\overline{\d_R} \phi_d^\dagger \chi\b_{L})\,(\overline{\d_L}  \chi\b_{R} )$,...]. 
Class-2 only contributes
to $B_{s}-\bar B_{s}$ mixing, up to $m_d/m_s$ corrections.
Taking into account that $SU(3)_F$ breaking in the bag parameters of the $B_{s}-\bar B_{s}$ vs.
$B_{d}-\bar B_{d}$ mixing matrix elements
is only at the few percent level in lattice QCD~\cite{Becirevic:2001xt}, we conclude that 
class-1 yields the \emph{same weak phase shift in $B_{d}-\bar B_{d}$ and 
$B_{s}-\bar B_{s}$ mixing relative to the SM}.  
The class-1 contribution would dominate if $\Lgmfv$ 
is comparable for all the operators.  For example, 
in the limit of equal Wilson coefficients $\kappa_i /\Lgmfv^2$,
the class-2 contribution to $B_{s}-\bar B_{s}$ mixing would be $\approx  5\%$ of class-1.
The maximal allowed magnitude 
of CPV in the $B_d$ system is smaller 
than roughly 20\%. Quantitatively, for $\Im\,\kappa_i \approx 1$, this 
corresponds to $\Lgmfv \approx 18$ TeV for the leading class-1 operator, 
which applies to the $B_s$ system as well. Thus, sizable CPV in the $B_s$ system would require 
class-2 contributions, with $O(1)$ CPV corresponding to $\Lgmfv \approx 1.5$ TeV for the leading class-2 operator.
Conversely, barring cancelations, \emph{within NLFMV models NP CPV in $B_{s}-\bar B_{s}$ mixing provides an upper bound on NP CPV in $B_{d}-\bar B_{d}$ mixing.}

For $2\to1$ transitions the new CPV phases come suppressed by powers of $m_{d,s}/m_b$.
All the $2\to 1$ bilinears in \eqref{eq:8}, \eqref{eq:10}, \eqref{eq:13}, \eqref{eq:13a}
are Hermitian with the exception of $\dbarL\chi\chi^\dagger \phi_d \d_R$. This provides the leading contribution to $\epsilon_K$ from a non-SM phase, coming from the operator ${\cal O}_{LR} = ( \dbarL \chi\chi^\dagger \phi_{d} \d_R)^2 $.   Its contribution is $\approx 2\% $ of the SM operator ${\cal O}_{LL} =(\dbarL\chi \chi^\dagger \d_L)^2$
for comparable Wilson coefficients $\kappa_{LR\,,LL}/\Lgmfv^2$.
For $\kappa_{LL} \,,\Im\,\kappa_{LR} \approx 1$, a new contribution to $\epsilon_K$ that 
is 50\% of the measured value 
would correspond to $\Lgmfv \approx 5$ TeV for ${\cal O}_{LL}$ and  $\Lgmfv \approx 0.8$ TeV for ${\cal O}_{LR}$.

Note that the above new CPV effects can only be sizable in the large $\tan \beta$ limit. They arise from non-Hermitian operators (such as the second operator in \eqref{eq:a1}), 
and are therefore of higher order in the  $Y_d$ expansion. Whereas we have been working in the large $\tan \beta$ limit,  it is straightforward to incorporate the small $\tan \beta$ limit into our formalism.
In that case the flavor group is broken down to $U(2)_Q\times U(2)_u\times U(1)_3\times U(3)_d$ and the expansion in Eq.~\eqref{eq:3} no longer holds.  In particular, resummation over $y_b$ is not required.  Flavor violation is described by linearly expanding in the down
type Yukawa couplings, from which it follows that contributions proportional to the bottom Yukawa are further suppressed beyond the
SM CKM suppression.

{\bf Up quark sector.}
Finally we comment on the up sector.  We work in the up-quark mass basis in which 
$\hat \rho_u=\hat \rho_d=0$, $\rho_Q=-\chi /2$, $\phi_u=\diag(m_u, m_c)/m_t$, while the flavor violating spurions are
$\phi_d=V_{CKM}^{(2)\dagger}\diag(m_d, m_s)/m_b$ and $\chi= i (V_{ub},V_{cb})$. An important prediction of the NLMFV models for the up-sector is that the new contributions
are greatly enhanced for large $\tan\beta$. Consider top flavor changing neutral currents (FCNC).
Within the SM they are highly suppressed by a combination of
a loop factor, GIM and CKM suppression. This results in branching ratios $BR(t\to c X)\sim {\cal O}(10^{-12})$.
An example of a FCNC bilinear operator in NLMFV is
$\overline{\tilde u^{(2)}} \chi \t$ [in the LMFV limit it corresponds to $\bar c_{L} \big(Y_b Y_b^\dagger\big)_{23} t_L$ and $\bar c_{L} \big(Y_b Y_b^\dagger\big)_{23} y_t t_R$].
 Model independent analysis shows that such an operator can lead to 
 $BR(t\to cX)\sim {\cal O}(10^{-5})$~\cite{Fox:2007in}, which may be within the reach of the LHC.

Similar enhancements are expected for CPV in $D-\bar D$ mixing.
The relevant operators are $(\ubar_L \chi\chi^\dagger \u_L)^2$ and
$(\ubar_L \chi\chi^\dagger \u_L)(\ubar_L \phi_d\phi_d^\dagger \u_L)$. 
The resulting CP violation 
in mixing is estimated to be ${\rm arg}(M_{12}/\Gamma_{12})= O(5\%)\, (1~{\rm TeV} /\Lgmfv)^2\, (\sin2\gamma ,\sin\gamma)$, 
%(1.6 \cdot 10^{-11}{\rm MeV} /\Delta M_D ) $,
respectively, where $\gamma=\arg(-V_{ud}V^*_{ub}/V_{cd}V_{cb}^*)$. 
For $\Lgmfv \sim 1$ TeV this is four orders of magnitude greater than in the SM, and would be observable in the future.
Operators of the type $(\ubarL \chi\chi^\dagger \phi_{u} \u_R)^2$ can contain a new CPV phase, but 
$(m_c/m_t )^2$ suppression renders them negligible by comparison. 
Unfortunately, experimental tests of MFV are generally very difficult in rare charm decays due to dominance of 
long-distance SM effects. 

{\bf Concluding remarks.}
Above we focused on the formalism and low energy flavor violating observables.
However, useful information can also be extracted 
from flavor diagonal quantities such as the new physics mass spectra~\cite{Grossman:2007bd} or non universal couplings to new gauge bosons~\cite{Fitzpatrick:2007sa}. For example, let us assume that 
new scalar states are in the fundamental of $U(3)_u$ so that the mass matrix squared is in its adjoint as in supersymmetric models
(e.g. right-handed squarks, neglecting the mixing with left-handed squarks). Order one or larger splitting $\Delta m_{13}^2$ between the first two and the third
generation would signal the NLMFV limit. Further insight would be provided by the mass spliting between the first two generations, $\Delta m_{12}^2$. In LMFV
$\Delta m_{13}^2:\Delta m_{12}^2=m_t^2:m_c^2$, while in NLMFV this relation would receive large corrections from subleading expansions in the Yukawas.
Finally we point out that NLMFV differs from the next-to-MFV (NMFV)~\cite{NMFV} framework since the latter exhibits additional spurions at low energy.  

{\bf Acknowledgements.} We thank N.~Arkani Hamed, J. Kamenik, D.E.~Kaplan, A.~Kronfeld, and A. Weiler for useful discussions, we also thank
 Y.~Nir and especially S.~Ja\"ger for discussions and comments on the manuscript.
A.~K. is supported by DOE grant FG02-84-ER40153, G.~P. by NSF grant PHY-06353354 and the
Peter and Patricia Gruber Award, T.~V. by DOE grant DE-FG02-90ER40542.

\end{document}